\documentclass[aps,preprint]{revtex4}
\usepackage{amsfonts}
\usepackage{amsmath}
\usepackage{amssymb}
\usepackage{graphicx}

\input{tcilatex}

\begin{document}

\preprint{}
\title{Regularization of the Singular Inverse Square Potential in Quantum
Mechanics with a Minimal length}
\author{Djamil Bouaziz Universit\'{e} de Li\`{e}ge, Institut de Physique B5,
Sart Tilman 4000 Li\`{e}ge 1, Belgium; and Laboratory of Theoretical
Physics, University of Jijel, BP 98, Ouled Aissa, 18000 Jijel, Algeria}
\author{Michel Bawin }
\affiliation{Universit\'{e} de Li\`{e}ge, Institut de Physique B5, Sart Tilman 4000 Li%
\`{e}ge 1, Belgium}
\keywords{}

\begin{abstract}
We study the problem of the attractive inverse square potential in quantum
mechanics with a generalized uncertainty relation. Using the momentum
representation, we show that this potential is regular in this framework. We
solve analytically the s-wave bound states equation in terms of Heun's
functions. We discuss in detail the bound states spectrum for a specific
form of the generalized uncertainty relation. The minimal length may be
interpreted as characterizing the dimension of the system.
\end{abstract}

\endpage{}
\maketitle

\section{\protect\bigskip Introduction}

\bigskip It is well known that in quantum gravity and string theory, there
is a lower bound to the possible resolution of distances, i.e., a minimal
observable length on the scale of the Planck length of $10^{-35}$ m. This
minimal length may be introduced as an additional uncertainty in position
measurements, so that the standard Heisenberg uncertainty relation becomes : 
$\left( \Delta X\right) \left( \Delta P\right) \geq\frac{\hbar}{2}%
[1+\beta\left( \Delta P\right) ^{2}+...]$, where $\beta$ is a small positive
parameter \cite{garay,amati,magiore}. It is clear that in this new relation, 
$\left( \Delta X\right) $ is always larger than $\left( \Delta X\right)
_{\min}=\hbar\sqrt{\beta}$. It was shown in Refs. \cite{k1,k7,k11,k2} that
the introduction of specific corrections to the usual canonical commutation
relations between position and momentum operators imply this new generalized
uncertainty relation in a natural way. This formalism, based on a
noncommutative Heisenberg algebra, together with the new concepts it
implies, has been discussed in one and more dimensions \cite{k1}. Quantum
field theory (QFT) has also been reformulated within this framework, and it
has been shown, in particular that, this minimal length may regularize
unwanted divergencies \cite{k4,k3}.

In addition to its importance in QFT, a minimal length may have a great
interest in nonrelativistic or relativistic quantum mechanics. Indeed, it
has been argued \cite{k7,smear} that this length may be viewed as an
intrinsic scale characterizing the system under study. Consequently, the
formalism based on these deformed commutation relations may provide a new
model for an effective description of complex systems such as
quasiparticules and various collectives excitations in solids, or composite
particules such as nucleons, nuclei, and molecules \cite{k7}. Various topics
were studied over the last ten years, in connection with this formalism: the
spectrum of the hydrogen atom has been obtained perturbatively in coordinate
space by several authors \cite{brau,ben,mmm,mm}, whereas its momentum space
treatment was done in Ref. \cite{akhoury}. The authors found an upper bound
of about $0.1$ fm for the minimal length by exploiting the experimental data
from precision hydrogen spectroscopy (the Lamb shift). The harmonic
oscillator potential has also been solved exactly in arbitrary dimensions 
\cite{chang} and perturbatively \cite{k1,k7,brau}. In Ref. \cite{chang}, an
upper bound for the minimal lengh has been calculated by confronting
theoretical results to precision measurement of electrons trapped in a
strong magnetic field; it is of the same order of magnitude as the result
obtained in the hydrogen atom problem. The influence of the minimal length
on the Casimir energy between two parallel plates has also been examined 
\cite{sab5,kh2}. The problem of a charged particle of spin one-half moving
in a constant magnetic field has been treated within the minimal length
formalism, and the thermal properties of the system at high temperatures
have been investigated \cite{kh3}. The minimal length was introduced in the
Dirac equation in Ref. \cite{kh1}, where a one-dimensional Dirac oscillator
has been solved exactly; in three dimensions, this problem has been solved
using supersymmetric quantum mechanics \cite{quesne}. Finally, the
modifications of the gyromagnetic moment of electrons and muons due to the
minimal length have been discussed in Ref. \cite{sab1}. For a review of
different approaches of theories with a minimal length scale and the
relation between them, we refer the reader to Ref. \cite{sab2}.

In this paper, we study the effect of a minimal length in nonrelativistic
quantum mechanics with a potential $V(R)$ of the form $V(R)=-\alpha/R^{2}$
with $2m\alpha/\hbar^{2}>1/4$ ($m$ is the particle mass). Such a potential
is singular when used in conjunction with the usual Schr\"{o}dinger
equation. Specifically, the condition of square integrability of the wave
function does not lead to an orthogonal set of eigenfunctions with their
corresponding eigenvalues \cite{case,perelo}. This is due to the fact that
the Hamiltonian operator is not self-adjoint \cite{metz}; to cure this
illness, we must define self-adjoint extensions of the Hamiltonian or
equivalentLy require othogonality of the wave functions \cite{case}.
However, the obtained spectrum is a peculiar one, as the energy eigenvalue
may take values from $0$ to $-\infty$, so that there is no finite ground
state. Landau and Lifshitz associate the occurrence of this infinite bound
state to the classical fall to the center of the particle \cite{landau}. In
addition to this fundamental problem, the expression of the energy spectrum
depends on an arbitrary phase parameter, coming from restoring the
self-adjointness of the Hamiltonian. For a review of works concerning this
potential, we refer the reader to Refs. \cite{wiliam,memoire}.

From a physical point of view, the strongly attractive $1/R^{2}$ potential
is very interesting. Indeed, the problem of atoms interacting with a charged
wire, relevant to the fabrication of nanoscale atom optical devices, is
known to provide an experimental realization of an attractive $1/R^{2}$
potential \cite{johane,bawin1}. It is a fundamental (long range) part of the
potential describing dipole-bound anions in polar molecules \cite{bawin2},
and has some applications in black holes physics \cite{horacio}. Finally,
let us note that the Efimov effect in three-body systems \cite{efimov}
arises from the existence of a long range effective interaction\ $V(R)$ of
the form $V(R)\sim$ $c/R^{2}$ ($c$ some constant), where $R$ is built from
the relative distances between the three particles. Further interest in the
singular inverse square potential also arose from recent studies showing
that it provides a simple example of a renormalization group limit cycle in
nonrelativistic quantum mechanics\ \cite{beane,bawin3,braten}. We also
mention for completeness sake other works on the regularization and the
renormalization of this potential \cite{gupta,camblong,coon}.

In this work we study in detail how the introduction of a generalized
uncertainty relation regularizes the singular inverse square potential in
nonrelativistic quantum mechanics. We show, in particular, that the
\textquotedblleft elementary length\textquotedblright\ included in these
relations may be interpreted as an effective cutoff regularizing the
potential at large momenta. It follows that in this new framework the
existence of an elementary length regularizes the $1/R^{2}$ potential,
without introducing any arbitrary cutoff.

Our paper is organized as follows. In section 2, we study the attractive $%
1/R^{2}$ potential in ordinary quantum mechanics, using the momentum
representation. In section 3, we derive the corresponding equations in
quantum mechanics with a modified uncertainty relation. In section 4, within
the formalism of deformed Heisenberg algebra, we solve exactly the Schr\"{o}%
dinger equation and extract the energy spectrum. Some concluding remarks are
reported in the last section.

\section{Singular attractive $1/R^{2}$ potential in ordinary quantum
mechanics}

The singular attractive inverse square potential has been extensively
studied in the coordinate representation (see for instance \cite%
{case,perelo,george,scarf,wiliam}). In Ref. \cite{perelo}, the expression of
the momentum wave function was given as a Fourier transform of the wave
function in configuration space. We use here a simple method for dealing
with the attractive $1/R^{2}$ potential in momentum space, as first applied
to the hydrogen atom potential \cite{eugene}.

\subsection{Schr\"{o}dinger equation in momentum representation}

We write the Schr\"{o}dinger equation for a particle of mass $m$ in the
external potential $V(R)=-\alpha/R^{2}$, $\alpha>0$ in the form 
\begin{equation}
(R^{2}P^{2}-2m\alpha)\left. \left\vert \psi\right. \right\rangle
=2mER^{2}\left. \left\vert \psi\right. \right\rangle \text{,}  \label{1}
\end{equation}
where $\overrightarrow{R}$ and $\overrightarrow{P}$ are, respectively, the
position and momentum operators. In the momentum representation, the wave
function reads \cite{chang} 
\begin{equation*}
\psi(\vec{p})=Y_{lm}(\theta,\varphi)\psi(p)
\end{equation*}
Without loss of generality, we restrict ourselves to $s$ waves. One then has 
\begin{align*}
R^{2}\psi(p) & =-\hbar^{2}\left( \frac{\partial^{2}}{\partial p^{2}}+\frac{2%
}{p}\frac{\partial}{\partial p}\right) \psi(p), \\
P^{2}\psi(p) & =p^{2}\psi(p).
\end{align*}

From Eq. (\ref{1}), we obtain the following differential equation:%
\begin{equation}
\frac{d^{2}\psi}{dp^{2}}+\frac{2}{p}\left( \frac{3p^{2}+k^{2}}{p^{2}+k^{2}}%
\right) \frac{d\psi}{dp}+\left( \frac{6+2m\alpha/\hbar^{2}}{p^{2}+k^{2}}%
\right) \psi=0,  \label{2}
\end{equation}
where $k^{2}=-2mE$.

Introducing the dimensionless variable $y$, defined by 
\begin{equation*}
y=-\frac{p^{2}}{k^{2}},
\end{equation*}
the Schr\"{o}dinger equation (\ref{2}) takes the following form:%
\begin{equation}
y(1-y)\frac{d^{2}\psi}{dy^{2}}+(\frac{3}{2}-\frac{7}{2}y)\frac{d\psi}{dy}-%
\frac{1}{2}(3+\frac{m\alpha}{\hbar^{2}})\psi=0.  \label{3}
\end{equation}

This equation is in the form of a hypergeometric equation \cite{abramo} as
follow:%
\begin{equation*}
y(1-y)\frac{d^{2}\psi}{dy^{2}}+\left[ c-(a+b+1)y\right] \frac{d\psi}{dy}%
-ab\psi=0,
\end{equation*}
with the parameters 
\begin{align}
a & =\frac{5}{4}+\frac{i}{2}\nu,  \notag \\
b & =\frac{5}{4}-\frac{i}{2}\nu,  \label{4} \\
c & =\frac{3}{2},\text{ where }\nu=\sqrt{\frac{2m\alpha}{\hbar^{2}}-1/4} 
\notag
\end{align}

The solution to Eq. (\ref{3}) finite for $p=0$ is \cite{abramo} 
\begin{equation}
\psi(p)=AF(a,b,c;-\frac{p^{2}}{k^{2}}),  \label{5}
\end{equation}
where $A$ is a normalization constant. This solution was obtained in Ref. 
\cite{perelo} by taking the Fourier transform of the configuration space
wave function $\psi(r)$, with%
\begin{equation*}
\psi(r)=Ar^{-1/2}K_{i\nu}(kr)
\end{equation*}
where $K_{i\nu}$ is the modified Bessel function.

Let us now examine the asymptotic behavior of solution (\ref{5}) in the
vicinity of $p=0$ and $p\rightarrow\infty$. For $p=0$, one has $\psi(p)=$
finite constant, as $F(a,b,c;y)\underset{y\ll1}{\approx}1$; so, it is
quadratically integrable at the origin. In the limit $p\rightarrow\infty$,
by means of the transformation \cite{abramo}\ 
\begin{align}
F(a,b,c;y) & =\frac{\Gamma(c)\Gamma(b-a)}{\Gamma(b)\Gamma(c-a)}%
(-y)^{-a}F(a,1-c+a,1-b+a;\frac{1}{y})+  \notag \\
& \frac{\Gamma(c)\Gamma(a-b)}{\Gamma(a)\Gamma(c-b)}(-y)^{-b}F(b,1-c+b,1-a+b;%
\frac{1}{y}),  \label{7}
\end{align}
the wave function (\ref{5}) is written as 
\begin{align}
\psi(p) & =\frac{\Gamma(3/2)\Gamma(-i\nu)}{\Gamma(5/4-\frac{i\nu}{2}%
)\Gamma(1/4-\frac{i\nu}{2})}\left( \frac{p}{k}\right) ^{-\frac{5}{2}-i\nu
}F(a,1-c+a,1-b+a;-\frac{k^{2}}{p^{2}})  \notag \\
& +\frac{\Gamma(3/2)\Gamma(i\nu)}{\Gamma(5/4+\frac{i\nu}{2})\Gamma (1/4+%
\frac{i\nu}{2})}\left( \frac{p}{k}\right) ^{-\frac{5}{2}+i\nu
}F(b,1-c+b,1-a+b;-\frac{k^{2}}{p^{2}}).  \label{8}
\end{align}

Then the behavior of $\psi(p)$ at infinity is of the form 
\begin{equation}
\psi(p)\underset{p\rightarrow\infty}{\sim}p^{-\frac{5}{2}}\left( Ap^{-i\nu
}+Bp^{+i\nu}\right) ,  \label{9}
\end{equation}
where $A$ and $B$ are complex constants.

Solution (\ref{8}) is a linear combination of two solutions that behave in
the same manner at infinity and, both of them, are quadratically integrable.
Usually the integrability condition suffices to distinguish between the two
independent solutions, but this is not the case here. From Eq. (\ref{9}),
one can see that the wave function depends on an arbitrary phase $\varphi$
as : $\psi(p)\underset{p\rightarrow\infty}{\sim}p^{-\frac{5}{2}}\cos\left(
\nu\ln p+\varphi\right) $, for real $\psi(p)$, and then it has an infinite
number of oscillations as $p\rightarrow\infty$. As was expected, the
oscillating behavior of $\psi(p)$ at infinity is analogous to the
oscillating behavior of the configuration space wave function $\psi(r)$ in
the neighborhood of the origin (see, for example, Ref. \cite{case}).

\subsection{Integral equation}

For later comparison with the solution of the Schr\"{o}dinger equation with
a minimal length, we derive now an integral equation equivalent to Eq. (\ref%
{2}). Let \ us observe that Eq. (\ref{1}) can be written in the form 
\begin{equation*}
\left[ L+g(p)\right] \varphi(p)=0,
\end{equation*}
where 
\begin{align*}
\varphi(p) & =(p^{2}+k^{2})\psi(p), \\
g(p) & =\frac{2m\alpha}{\hbar^{2}}\frac{p^{2}}{(p^{2}+k^{2})},
\end{align*}
and $L$ is the self-adjoint operator 
\begin{equation}
L=-\frac{p^{2}}{\hbar^{2}}R^{2}=\frac{d}{dp}\left( p^{2}\frac{d}{dp}\right) .
\label{9b}
\end{equation}
Then $\varphi(p)$, satisfying the boundary conditions $\varphi(p)=$ constant
and $\varphi(\infty)=0$, is given by \cite{fesh1}\ 
\begin{equation}
\varphi(p)=\int_{0}^{\infty}G(p,p^{^{\prime}})g(p^{^{\prime}})\varphi
(p^{^{\prime}})dp^{^{\prime}}.  \label{9c}
\end{equation}
The Green function $G(p,p^{^{\prime}})$ is then given by 
\begin{equation}
G(p,p^{^{\prime}})=\frac{\theta(p-p^{^{\prime}})}{p}+\frac{%
\theta(p^{^{\prime }}-p)}{p^{^{\prime}}},  \label{9e}
\end{equation}
and the integral equation satisfied by the wave function $\psi(p)$ is 
\begin{equation}
(p^{2}+k^{2})\psi(p)=\frac{2m\alpha}{\hbar^{2}}\int_{0}^{\infty}p^{^{%
\prime}2}\psi(p^{^{\prime}})G(p,p^{^{\prime}})dp^{^{\prime}}.  \label{9f}
\end{equation}
This equation can also be obtained by calculating the Fourier transform of
the potential and inserting it in the s-wave integral Schr\"{o}dinger
equation and then integrating over the angles \cite{hamer}\ .

Note that putting $\psi(p)\sim p^{s}$ in Eq. (\ref{9f}), we get :%
\begin{equation*}
p^{s+2}\underset{p\sim\infty}{=}\frac{2m\alpha}{\hbar^{2}}\left\{ \frac{1}{p}%
\int^{p\sim\infty}p^{^{\prime}s+2}dp^{^{\prime}}+\int_{p\sim\infty
}p^{^{\prime}s+1}dp^{^{\prime}}\right\}
\end{equation*}
After integration we get the characteristic equation 
\begin{equation*}
1=\frac{2m\alpha}{\hbar^{2}}\left[ \frac{1}{s+3}-\frac{1}{s+2}\right] ,
\end{equation*}
which has two roots $s=-\frac{5}{2}\pm i\nu$, corresponding to the two
solutions (\ref{9}).

This is the momentum space illustration of the singular nature of the
potential $-\alpha/R^{2}$ : Eq. (\ref{9f}) has square integrable solutions
for any value of $k^{2}>0$.

\subsection{Energy spectrum}

For completeness sake, we now show, following \cite{fesh}, how a spectrum
can be obtained by requiring the functions $\psi(p)$ to be mutually
orthogonal.

\subsubsection{Orthogonality of the eigenfunctions}

Let us consider two eigenfunctions $\psi_{1}(p)$ and $\psi_{2}(p)$
corresponding, respectively, to the eigenvalues $k_{1}$ and $k_{2}$. The
scalar product between these two functions reads 
\begin{equation}
\left\langle \psi_{1}\left\vert \psi_{2}\right. \right\rangle
=A_{1}A_{2}^{\ast}\int_{0}^{\infty}p^{2}dpF(a,b,c;-\frac{p^{2}}{k_{1}^{2}}%
)F(a,b,c;-\frac{p^{2}}{k_{2}^{2}}).  \label{10}
\end{equation}

Introducing the change of variable $x=p^{2}$ and using the formula \cite%
{russe} 
\begin{align*}
& \int_{0}^{\infty}x^{c-1}F(a,b,c;-\sigma
x)F(a^{^{\prime}},b^{^{\prime}},c;-\omega x)dx \\
& =\sigma^{-a}\omega^{a-c}\dfrac{\left[ \Gamma(c)\right] ^{2}\Gamma(a+a^{^{%
\prime}}-c)\Gamma(a+b^{^{\prime}}-c)\Gamma(a^{^{\prime}}+b-c)\Gamma(b+b^{^{%
\prime}}-c)}{\Gamma(a)\Gamma(b)\Gamma(a^{^{\prime}})\Gamma(b^{^{\prime}})%
\Gamma(a+a^{^{\prime}}+b+b^{^{\prime}}-2c)} \\
& \times F(a+a^{^{\prime}}-c,a+b-c,a+a^{^{\prime}}+b+b^{^{\prime}}-2c;1-%
\frac{\omega}{\sigma}),
\end{align*}%
\begin{align*}
\func{Re}c\text{,}\func{Re}(a+a^{^{\prime}}-c)\text{, }\func{Re}%
(a+b^{^{\prime}}-c)\text{, }\func{Re}(a^{^{\prime}}+b-c)\text{, }\func{Re}%
(b+b^{^{\prime}}-c) & >0 \\
\left\vert \arg\sigma\right\vert ,\left\vert \arg\omega\right\vert & <\pi,
\end{align*}
we obtain 
\begin{equation*}
\left\langle \psi_{1}\left\vert \psi_{2}\right. \right\rangle =\Omega\left( 
\frac{k_{1}}{k_{2}}\right) ^{i\nu}F(1+i\nu,1,2;1-\frac{k_{1}^{2}}{k_{2}^{2}}%
),
\end{equation*}
where 
\begin{equation*}
\Omega=\frac{1}{2}A_{1}A_{2}^{\ast}k_{1}^{\frac{5}{2}}k_{2}^{\frac{1}{2}}%
\dfrac{\left[ \Gamma(\frac{3}{2})\right] ^{2}\left[ \Gamma(1)\right]
^{2}\Gamma(1+i\nu)\Gamma(1-i\nu)}{\Gamma(2)\left[ \Gamma(\frac{5}{4}+i\frac{%
\nu}{2})\right] ^{2}\left[ \Gamma(\frac{5}{4}-i\frac{\nu}{2})\right] ^{2}}.
\end{equation*}
Using the formula \cite{abramo} 
\begin{align*}
F(a,b,c;z) & =\frac{1}{b-1-(c-a-1)z}\left[ (b-c)F(a,b-1,c;z)\right. \\
& \left. +(c-1)(1-z)F(a,b,c-1;z)\right] ,
\end{align*}
and 
\begin{equation*}
F(a,b,b;z)=(1-z)^{-a}\text{, \ }F(0,b,c;z)=F(a,0,c;z)=1,
\end{equation*}
we get, finally, the following expression for the scalar product:%
\begin{align}
\left\langle \psi_{1}\left\vert \psi_{2}\right. \right\rangle & =\frac{\Omega%
}{i\nu(\frac{k_{1}^{2}}{k_{2}^{2}}-1)}\left[ \left( \frac {k_{1}}{k_{2}}%
\right) ^{+i\nu}-\left( \frac{k_{1}}{k_{2}}\right) ^{-i\nu }\right]  \notag
\\
& =\frac{2\Omega}{\nu(\frac{k_{1}^{2}}{k_{2}^{2}}-1)}\sin\left[ \nu\ln (%
\frac{k_{1}}{k_{2}})\right] .  \label{11}
\end{align}

\bigskip It is clear that $\psi_{1}$ and $\psi_{2}$ are orthogonal, if the
following condition is satisfied :%
\begin{equation}
\nu\ln(\frac{k_{1}}{k_{2}})=n\pi,\text{ \ \ \ \ }n=0,\pm1,...\text{ .}
\label{12}
\end{equation}
This condition leads to the following discrete spectrum :%
\begin{equation}
E_{n}=E_{1}\exp[-\frac{2n\pi}{\nu}],\text{ \ \ \ }n=0,\pm1,...\text{ .}
\label{13}
\end{equation}

It is the same result as obtained in coordinate space by Case \cite{case}\ .
Thus a requirement that the state functions for bound states, for $%
2m\alpha/\hbar^{2}>1/4$, be a mutually orthogonal set imposes a quantization
of energy. It does not uniquely fix the levels, but it fixes the levels
relative to one another. If we fix $E_{1}$, then the bound levels extend to $%
-\infty$ and have an accumulation point at zero energy \cite{fesh}.

Now we show that the energy spectrum can be obtained by introducing a
momentum space cutoff $\Lambda\gg k$ with the boundary condition $%
\psi(\Lambda)=0$. We note that this regularization procedure was used in
Refs. \cite{gupta,camblong,coon}, in coordinate space. This regularization
is equivalent to replacing the potential at short distances with an
infinitely repulsive barrier.

\subsubsection{Regularization by an ultraviolet cutoff}

Let us go back to the wave function (\ref{8}), by writing the boundary
condition $\psi(\Lambda)\underset{\Lambda\gg k}{=}0$.

Bearing in mind that $F(a,b,c;y)\underset{y\ll1}{\approx}1$, we obtain 
\begin{equation}
\left( \frac{\Lambda}{k}\right) ^{-\frac{5}{2}-i\nu}\exp[-i\arg(A)]+\left( 
\frac{\Lambda}{k}\right) ^{-\frac{5}{2}+i\nu}\exp[i\arg(A)]=0,  \label{14}
\end{equation}
where 
\begin{equation*}
A\equiv\frac{\Gamma(i\nu)}{\Gamma(5/4+\frac{i\nu}{2})\Gamma(1/4+\frac{i\nu}{2%
})}=\left\vert A\right\vert \exp[i\arg(A)],
\end{equation*}
Eq. (\ref{14}) can be written as 
\begin{equation}
\cos[\arg(A)+\nu\ln(\frac{\Lambda}{k})]=0,  \label{15}
\end{equation}
which gives the following bound states : 
\begin{align}
E_{n} & =-\frac{k^{2}}{2m}=-\frac{\Lambda^{2}}{2m}\exp\frac{2}{\nu}\left[
\arg(A)-(n+\frac{1}{2})\pi\right] ,  \notag \\
n & =0,+1,+2,...\text{ .}  \label{16}
\end{align}

Consequently, this regularization leads to a quantized energy spectrum,
which now possesses a finite ground state for the singular attractive $%
1/R^{2}$ potential.

\section{Quantum mechanics with a generalized uncertainty relation}

Let us consider the following modified commutation relation between the
position and momentum operators:%
\begin{equation}
\lbrack\widehat{X},\widehat{P}]=i\hbar\left( 1+\beta\widehat{P}^{2}\right) ,%
\text{ \ }\beta>0  \label{17}
\end{equation}
This commutation relation leads to the generalized uncertainty relation \cite%
{k1} 
\begin{equation}
\left( \Delta X\right) \left( \Delta P\right) \geq\frac{\hbar}{2}\left(
1+\beta\left( \Delta P\right) ^{2}+\beta\left\langle \widehat{P}%
\right\rangle ^{2}\right) ,  \label{18}
\end{equation}
which implies a lower bound for $\Delta X$ or a minimal length, given by 
\begin{equation}
\left( \Delta X\right) _{\min}=\hbar\sqrt{\beta}  \label{19}
\end{equation}
The striking feature of Eq. (\ref{18}) is the UV/IR mixing: when $\Delta P$
is large (UV), $\Delta X$ is proportional to $\Delta P$ and, therefore, is
also large (IR). This phenomenon is said to be necessary to understand the
cosmological constant problem or the observable implications of short
distance physics on inflationary cosmology; it has appeared in several
contexts for example, in noncommutative field theory \cite{sandore}. Another
fundamental consequence of the minimal length is the loss of localization in
coordinates space, so that, momentum space is more convenient in order to
solve any eigenvalue problem.

An explicit form for $\widehat{X}$ and $\widehat{P}$ satisfying Eq. (\ref{17}%
) is given by 
\begin{equation}
\begin{array}{c}
\widehat{X}=i\hbar\lbrack(1+\beta p^{2})\dfrac{\partial}{\partial p}+\gamma
p], \\ 
\widehat{P}=p,%
\end{array}
\label{20}
\end{equation}
where a constant $\gamma$ does not affect the observables quantities; it
determines only the weight function in the definition of the scalar product 
\cite{chang} as follow: 
\begin{equation}
\langle\varphi\mid\psi\rangle=\int_{-\infty}^{+\infty}\frac{dp}{\left(
1+\beta p^{2}\right) ^{1-\frac{\gamma}{\beta}}}\varphi^{\ast}(p)\psi(p).
\label{21}
\end{equation}

A generalization of Eq. (\ref{17}) to $D$\ dimensions is \cite%
{k1,k7,chang,sandore} : 
\begin{equation}
\lbrack\widehat{X}_{i},\widehat{P}_{j}]=i\hbar\lbrack(1+\beta\widehat{P}%
^{2})\delta_{ij}+\beta^{^{\prime}}\widehat{P}_{i}\widehat{P}_{j}],\text{ \ \ 
}(\beta,\beta^{^{\prime}})>0.  \label{22}
\end{equation}
If we assume that 
\begin{equation}
\lbrack\widehat{P}_{i},\widehat{P}_{j}]=0,  \label{23}
\end{equation}
then the Jacobi identity determines the commutation relations among the
coordinates $\widehat{X}_{i}$ as 
\begin{equation}
\lbrack\widehat{X}_{i},\widehat{X}_{j}]=i\hbar\frac{2\beta-\beta^{^{%
\prime}}+\beta(2\beta+\beta^{^{\prime}})\widehat{P}^{2}}{1+\beta\widehat{P}%
^{2}}\left( \widehat{P}_{i}\widehat{X}_{j}-\widehat{P}_{j}\widehat{X}%
_{i}\right) .  \label{24}
\end{equation}

\bigskip The generalized uncertainty relation implied by, Eq. (\ref{22}) is 
\begin{equation}
\left( \Delta X_{i}\right) \left( \Delta P_{i}\right) \geq\frac{\hbar}{2}%
\left( 1+\beta\tsum \limits_{j=1}^{D}[\left( \Delta P_{j}\right)
^{2}+\left\langle \widehat{P}_{j}\right\rangle
^{2}]+\beta^{^{\prime}}[\left( \Delta P_{i}\right) ^{2}+\left\langle 
\widehat{P}_{i}\right\rangle ^{2}]\right) .  \label{25}
\end{equation}
This relation leads to a lower bound of $\Delta X_{i}$, given by 
\begin{equation}
\left( \Delta X_{i}\right) _{\min}=\hbar\sqrt{\left( D\beta+\beta
^{^{\prime}}\right) },\text{ \ \ }\forall i.  \label{26}
\end{equation}

In the momentum representation, the following realization satisfies the
above commutation relations: 
\begin{equation}
\begin{array}{c}
\widehat{X}_{i}=i\hbar\left( (1+\beta p^{2})\dfrac{\partial}{\partial p_{i}}%
+\beta^{^{\prime}}p_{i}p_{j}\dfrac{\partial}{\partial p_{j}}+\gamma
p_{i}\right) , \\ 
\widehat{P}_{i}=p_{i}.%
\end{array}
\label{27}
\end{equation}
As in one dimension, the arbitrary constant $\gamma$ does not affect the
observable quantities, its choice determines the weight factor in the
definition of the scalar product as follow: 
\begin{align}
\left\langle \varphi\left\vert \psi\right. \right\rangle & =\int\frac {d^{D}p%
}{\left[ 1+\left( \beta+\beta^{^{\prime}}\right) p^{2}\right] ^{1-\alpha}}%
\varphi^{\ast}(p)\psi(p),  \notag \\
\alpha & =\frac{\gamma-\beta^{^{\prime}}\left( \frac{D-1}{2}\right) }{%
\beta+\beta^{^{\prime}}}.  \label{28}
\end{align}

\section{Singular attractive $1/R^{2}$ potential in quantum mechanics with a
generalized uncertainty relation}

\subsection{The Schr\"{o}dinger equation}

We proceed, as in Sec. II, by writing the Schr\"{o}dinger equation, for a
particle of mass $m$ in the external potential $V(R)=-\alpha/R^{2}$, $%
\alpha>0$, in the form 
\begin{equation}
(R^{2}P^{2}-2m\alpha)\left. \left\vert \psi\right. \right\rangle
=2mER^{2}\left. \left\vert \psi\right. \right\rangle .  \label{29}
\end{equation}
Restricting ourselves to the $l=0$ wave function and using Eq. (\ref{27})
with $\gamma=0$, we obtain the following expression for $R^{2}\equiv\tsum
\limits_{i=1}^{3}X_{i}X_{i}$:%
\begin{equation}
R^{2}=\left( i\hbar\right) ^{2}\left\{ \left[ 1+(\beta+\beta^{^{%
\prime}})p^{2}\right] ^{2}\frac{d^{2}}{dp^{2}}+\left[ 1+(\beta+\beta^{^{%
\prime}})p^{2}\right] \left[ 2(2\beta+\beta^{^{\prime}})p+\frac{2}{p}\right] 
\frac{d}{dp}\right\} .  \label{30}
\end{equation}

From Eqs. (\ref{29}) and Eq. (\ref{30}) the Schr\"{o}dinger equation for the 
$-\alpha/R^{2}$\ potential in the presence of a minimal length takes the
form 
\begin{equation*}
\frac{d^{2}\psi(p)}{dp^{2}}+\frac{2}{p}\left\{ 4\left[ \frac{p^{2}-mE}{%
p^{2}-2mE}\right] -\frac{1+\beta^{^{\prime}}p^{2}}{1+(\beta+\beta^{^{\prime
}})p^{2}}\right\} \frac{d\psi(p)}{dp}+
\end{equation*}%
\begin{equation}
+\left\{ \frac{6+(10\beta+6\beta^{^{\prime}})p^{2}}{\left[ 1+(\beta
+\beta^{^{\prime}})p^{2}\right] }+\frac{2m\alpha/\hbar^{2}}{\left[
1+(\beta+\beta^{^{\prime}})p^{2}\right] ^{2}}\right\} \frac{\psi(p)}{%
(p^{2}-2mE)}=0.  \label{31}
\end{equation}

In the case $\beta=\beta^{^{\prime}}=0$, this equation reduces to Eq. (\ref%
{2}) of ordinary quantum mechanics.

We can again transform Eq. (\ref{31}) to an integral equation. We write Eq. (%
\ref{29}) in the form 
\begin{equation}
R^{2}\varphi(p)=2m\alpha\psi(p),  \label{31a}
\end{equation}
where 
\begin{equation*}
\varphi(p)=(p^{2}-2mE)\psi(p).
\end{equation*}
Then $R^{2}$ can be written as : 
\begin{equation*}
R^{2}=-\hbar^{2}p^{-2}\left[ 1+(\beta+\beta^{^{\prime}})p^{2}\right] ^{2-%
\frac{\beta}{\beta+\beta^{^{\prime}}}}\widetilde{L},
\end{equation*}
where $\widetilde{L}$ is the following self-adjoint operator: 
\begin{equation}
\widetilde{L}=\frac{d}{dp}\left( K(p)\frac{d}{dp}\right) ,  \label{31aa}
\end{equation}
with 
\begin{equation*}
K(p)=p^{2}\left[ 1+(\beta+\beta^{^{\prime}})p^{2}\right] ^{\frac{\beta }{%
\beta+\beta^{^{\prime}}}}.
\end{equation*}
Eq. (\ref{31a}) is then transformed to the following nonhomogeneous
Sturm-Liouville equation:%
\begin{equation}
\left[ \widetilde{L}+g(p)\right] \varphi(p)=0,  \label{31b}
\end{equation}
where 
\begin{equation}
g(p)=\frac{2m\alpha}{\hbar^{2}}\frac{p^{2}}{p^{2}-2mE}\left[ 1+(\beta
+\beta^{^{\prime}})p^{2}\right] ^{\frac{\beta}{\beta+\beta^{^{\prime}}}-2}.
\label{31c}
\end{equation}
Then $\varphi(p)$ is given by the integral \cite{fesh1} 
\begin{align}
\varphi(p) &
=\int_{0}^{\infty}G(p,p^{^{\prime}})g(p^{^{\prime}})\varphi(p^{^{%
\prime}})dp^{^{\prime}}+\left[ \varphi(0)K(0)\frac {dG(p,p^{^{\prime}})}{%
dp^{^{\prime}}}\right] _{p^{^{\prime}}=0}  \notag \\
& -\left[ \varphi(\infty)K(\infty)\frac{dG(p,p^{^{\prime}})}{dp^{^{\prime}}}%
\right] _{p^{^{\prime}}=\infty}.  \label{31d}
\end{align}
$G(p,p^{^{\prime}})$ is the corresponding Green's function.

In order to have a homogeneous integral equation in the form of an
eigeinvalue problem 
\begin{equation}
\varphi(p)=\int_{0}^{\infty}G(p,p^{^{\prime}})g(p^{^{\prime}})\varphi
(p^{^{\prime}})dp^{^{\prime}},  \label{31da}
\end{equation}
$\varphi(p)$ must vanish at infinity. The wave function $\psi(p)$ is then
required to satisfy the boundary condition 
\begin{equation}
p^{2}\psi(p)\underset{p\rightarrow\infty}{=}0.  \label{31dd}
\end{equation}

The explicit form of $G(p,p^{^{\prime}})$, using the boundary conditions (%
\ref{31dd}), and $\psi(0)=constant$, is found to be 
\begin{equation*}
G(p,p^{^{\prime}})=\left\{ 
\begin{array}{c}
\frac{1}{p}F(-\frac{1}{2},\frac{\beta}{\beta+\beta^{^{\prime}}},\frac{1}{2}%
;-[\beta+\beta^{^{\prime}}]p^{2})-C,\text{ \ \ }p>p^{^{\prime}}, \\ 
\frac{1}{p^{^{\prime}}}F(-\frac{1}{2},\frac{\beta}{\beta+\beta^{^{\prime}}},%
\frac{1}{2};-[\beta+\beta^{^{\prime}}]p^{^{\prime}2})-C,\text{ \ \ }%
p<p^{^{\prime}},%
\end{array}
\right.
\end{equation*}
where $C$ is the constant 
\begin{equation}
C=\dfrac{(\beta+\beta^{^{\prime}})^{\frac{1}{2}}\Gamma(\frac{1}{2})\Gamma(%
\frac{1}{2}+\frac{\beta}{\beta+\beta^{^{\prime}}})}{\Gamma (1)\Gamma(\frac{%
\beta}{\beta+\beta^{^{\prime}}})}.  \label{31ee}
\end{equation}

Finally, the integral equation satisfied by the wave function $\psi(p)$ is 
\begin{equation}
(p^{2}-2mE)\psi(p)=\frac{2m\alpha}{\hbar^{2}}\int_{0}^{\infty}p^{^{\prime}2}%
\left[ 1+(\beta+\beta^{^{\prime}})p^{^{\prime}2}\right] ^{\frac{\beta }{%
\beta+\beta^{^{\prime}}}-2}G(p,p^{^{\prime}})\psi(p^{^{\prime}})dp^{^{%
\prime}}.  \label{31f}
\end{equation}

In the limit $\beta=\beta^{^{\prime}}=0$, Eq. (\ref{31f}) reduces to Eq. (%
\ref{9f}) of ordinary quantum mechanics.

\bigskip Let us return now to the differential equation (\ref{31}); by
introducing the dimensionless variable $z$, defined as 
\begin{equation}
z=\frac{(\beta+\beta^{^{\prime}})p^{2}-1}{(\beta+\beta^{^{\prime}})p^{2}+1},
\label{32}
\end{equation}
which varies from $-1$ to $+1$, and using the following notations:%
\begin{align}
\omega_{1} & =\beta+\beta^{^{\prime}},\text{ \ }\omega_{2}=\beta
+2\beta^{^{\prime}},\text{ \ }\omega_{3}=2\beta+3\beta^{^{\prime}},  \notag
\\
\omega_{4} & =\frac{\beta}{\beta+\beta^{^{\prime}}},\text{ \ }%
\omega=-m(\beta+\beta^{^{\prime}})E,\text{ \ }\kappa=\frac{m\alpha}{2\hbar
^{2}}  \label{33}
\end{align}
we obtain the differential equation 
\begin{equation*}
(1-z^{2})\frac{d^{2}\psi}{dz^{2}}+\left[ 8\frac{(1+\omega)+(1-\omega )z}{%
(1+2\omega)+(1-2\omega)z}-\frac{1}{\omega_{1}}(\omega_{2}z+\omega _{3})%
\right] \frac{d\psi}{dz}
\end{equation*}%
\begin{equation}
+\left[ \dfrac{\kappa z^{2}+2(\omega_{4}-\kappa)z+(6+2\omega_{4}+\kappa )}{%
(-1+2\omega)z^{2}-4\omega z+(1+2\omega)}\right] \psi=0.  \label{34}
\end{equation}

To rewrite this equation in the form of a known differential equation, we
make the following transformation: 
\begin{equation}
\psi(z)=(1-z)^{\lambda}(1+z)^{\lambda^{^{\prime}}}f(z)  \label{35}
\end{equation}
where $\lambda$ and $\lambda^{^{\prime}}$ are arbitrary constants. Then, the
equation for $f(z)$ is 
\begin{equation*}
\frac{d^{2}f}{dz^{2}}+\left\{ \frac{-2\lambda}{(1-z)}+\frac{2\lambda
^{^{\prime}}}{(1+z)}+\frac{8\left[ (1+\omega)+(1-\omega)z\right] }{(1-z^{2})%
\left[ (1+2\omega)+(1-2\omega)z\right] }-\frac{(\omega_{2}z+\omega_{3}}{%
\omega_{1}(1-z^{2})}\right\} \frac{df}{dz}
\end{equation*}%
\begin{equation*}
+\left\{ \frac{\lambda(\lambda-1)}{(1-z)^{2}}+\frac{\lambda^{^{\prime}}(%
\lambda^{^{\prime}}-1)}{(1+z)^{2}}-\frac{8\lambda\left[ (1+\omega
)+(1-\omega)z\right] }{(1-z^{2})(1-z)\left[ (1+2\omega)+(1-2\omega)z\right] }%
+\right.
\end{equation*}%
\begin{equation*}
\frac{8\lambda^{^{\prime}}\left[ (1+\omega)+(1-\omega)z\right] }{%
(1-z^{2})(1+z)\left[ (1+2\omega)+(1-2\omega)z\right] }+\lambda\frac {%
(\omega_{2}z+\omega_{3})}{\omega_{1}(1-z)^{2}(1+z)}-\lambda^{^{\prime}}\frac{%
(\omega_{2}z+\omega_{3})}{\omega_{1}(1-z)(1+z)^{2}}
\end{equation*}%
\begin{equation}
\left. +\dfrac{\kappa z^{2}+2(\omega_{4}-\kappa)z+(6+2\omega_{4}+\kappa )}{%
(1-z^{2})\left[ (-1+2\omega)z^{2}-4\omega z+(1+2\omega)\right] }-\frac{%
2\lambda\lambda^{^{\prime}}}{(1-z^{2})}\right\} f=0.  \label{36}
\end{equation}
\ 

This equation constitutes our starting point for studying the attractive $1/R%
{{}^2}%
$ potential in quantum mechanics with a minimal length. We shall be
interested in the singularity structure of this equation and the effect of
the finite length. For this purpose, let us begin with the case $E=0$.

\subsection{Zero energy solution}

The simplicity of the zero energy Schr\"{o}dinger equation allows us to
investigate whether the \textquotedblright deformed" version of the $%
-\alpha/R^{2}$ potential in momentum space from Eq. (\ref{36}) remains
singular.

Let us rewrite Eq. (\ref{36}), in the case $\omega=0$ in the following form:%
\begin{equation*}
(1-z^{2})\frac{d^{2}f}{dz^{2}}+\left\{
(-2\lambda+2\lambda^{^{\prime}}+5+\omega_{4})-(2\lambda+2\lambda^{^{%
\prime}}+2-\omega_{4})z\right\} \frac{df}{dz}+
\end{equation*}%
\begin{equation*}
\frac{1}{(1-z^{2})}\left\{ (1+z)\left[ \lambda(\lambda-1)(1+z)-2\lambda
\lambda^{^{\prime}}(1-z)-\lambda(5+\omega_{4})+\lambda(2-\omega_{4})z+2%
\omega_{4}\right] \right.
\end{equation*}%
\begin{equation}
\left. +(1-z)\left[ \lambda^{^{\prime}}(\lambda^{^{\prime}}-1)(1-z)+\lambda
^{^{\prime}}(5+\omega_{4})-\lambda^{^{\prime}}(2-\omega_{4})z+\kappa (1-z)%
\right] +6\right\} f=0.  \label{37}
\end{equation}

We choose $\lambda$ and $\lambda^{^{\prime}}$ by requiring that the
coefficient \ of $f(z)$ in Eq. (\ref{37}) vanishes for $z=\pm1$; this leads
to the two equations for $\lambda$ and $\lambda^{^{\prime}}$ as follow:%
\begin{equation}
\begin{array}{c}
\lambda^{2}-(\frac{5}{2}+\omega_{4})\lambda+\frac{3}{2}+\omega_{4}=0, \\ 
\lambda^{^{\prime}2}+\frac{5}{2}\lambda^{^{\prime}}+\kappa+\frac{3}{2}=0,%
\end{array}
\label{38}
\end{equation}
The values of $\lambda$ and $\lambda^{^{\prime}}$ satisfying this system are 
\begin{align*}
\lambda & =1,\text{ }(\frac{3}{2}+\omega_{4}) \\
\lambda^{^{\prime}} & =(-\frac{5}{4}-i\frac{\nu}{2}),\text{ }(-\frac{5}{4}+i%
\frac{\nu}{2}),
\end{align*}
where $\nu=\sqrt{4\kappa-1/4}$. We note that there are four possible choices
concerning $(\lambda,\lambda^{^{\prime}})$ leading to the same solution of
the Schr\"{o}dinger equation. We select the set $(1,-\frac{5}{4}-i\frac{\nu}{%
2})$; so the transformation (\ref{35}) becomes 
\begin{equation}
\psi(z)=(1-z)(1+z)^{(-\frac{5}{4}-i\frac{\nu}{2})}f(z).  \label{39}
\end{equation}

By substituting $\lambda$ and $\lambda^{^{\prime}}$ with their values in Eq.
(\ref{37}), we obtain 
\begin{equation}
(1-z^{2})\frac{d^{2}f}{dz^{2}}+\left\{ (\frac{1}{2}+\omega_{4}-i\nu )-(\frac{%
3}{2}-\omega_{4}-i\nu)z\right\} \frac{df}{dz}+\left\{ (\frac{1}{8}-\frac{%
\omega_{4}}{4})+i\nu(\frac{1}{4}-\frac{\omega_{4}}{2})\right\} f=0.
\label{40}
\end{equation}

This equation is a second-order differential equation with three (regular)
singular points $z=1,-1,\infty$. Consequently, it may be written in a
canonical form of a hypergeometric equation, merely by transforming the
singular points to $z=0,1,\infty$. We can do this by means of the simple
following change of variable:%
\begin{equation}
\xi=\frac{z+1}{2}.  \label{41}
\end{equation}

Thus, Eq. (\ref{40}) becomes 
\begin{equation}
\xi(1-\xi)f^{^{\prime\prime}}(\xi)+\left[ c-(a+b+1)\xi\right]
f^{^{\prime}}(\xi)-abf(\xi)=0,  \label{42}
\end{equation}

\bigskip with the parameters 
\begin{align}
a & =\frac{1}{4}-\frac{\omega_{4}}{2}-\frac{\mu}{2}-i\frac{\nu}{2},  \notag
\\
b & =\frac{1}{4}-\frac{\omega_{4}}{2}+\frac{\mu}{2}-i\frac{\nu}{2},  \notag
\\
c & =1-i\nu\text{, \ \ \ }\nu=\sqrt{4\kappa-1/4},  \label{43} \\
\mu & =\sqrt{(\omega_{4}-1)^{2}-4\kappa}\text{, \ \ \ }\kappa=m\alpha
/2\hbar^{2}.  \notag
\end{align}

Equation (\ref{40}) is a hypergeometric equation which has, in the
neighborhood of $\xi=0$, the following two solutions \cite{abramo}: 
\begin{align}
f_{1}(\xi) & =F\left( a,b,c;\xi\right) ,  \label{44} \\
f_{2}(\xi) & =\xi^{1-c}F\left( a-c+1,b-c+1,2-c;\xi\right) ,  \label{45}
\end{align}
where $F\left( a,b,c;\xi\right) \equiv_{2}F_{1}\left( a,b,c;\xi\right) $ is
the hypergeometric function.

Finally, from Eq. (\ref{39}), we obtain two solutions $\psi_{1}(\xi)$ and $%
\psi_{2}(\xi)$, each solution being the complex conjugate of the other.
Thus, the general solution is 
\begin{equation}
\psi(\xi)=(1-\xi)\xi^{-\frac{5}{4}}\left[ A\xi^{-i\frac{\nu}{2}}F\left(
a,b,c;\xi\right) +B\xi^{i\frac{\nu}{2}}F\left( a-c+1,b-c+1,2-c;\xi\right) %
\right]  \label{46}
\end{equation}

In the particular case where $\omega_{4}=1/2$ ($\beta=\beta^{^{\prime}}$),
we have $\mu=i\nu$ and $b=0$. As $F\left( a,0,c;\xi\right) =1$, the wave
function $\psi(\xi)$ simplifies to :%
\begin{equation}
\psi_{\beta=\beta^{^{\prime}}}(\xi)=(1-\xi)\xi^{-\frac{5}{4}}\left[ A\xi^{-i%
\frac{\nu}{2}}+B\xi^{+i\frac{\nu}{2}}\right] .  \label{47}
\end{equation}

In the limit $\beta,\beta^{^{\prime}}\ll1$, one has $\xi\sim\omega_{1}p^{2}%
\ll1$, so that $1-\xi\sim1$ and $F\left( a,b,c;\xi\right) \sim1$.
Consequently, Eq. (\ref{46}) becomes 
\begin{equation}
\psi(p)\underset{\omega_{1}\ll1}{\sim}p^{-5/2}(Ap^{-i\nu}+Bp^{+i\nu}).
\label{48}
\end{equation}
This is exactly the zero energy solution of ordinary quantum mechanics,
which has the same form as the solution in the limit $p\rightarrow\infty$
[see Eq. (\ref{9})].

Solutions (\ref{46}) have the same behavior near $\xi=0$. This is not so,
however, for $p\rightarrow\infty$ ($\xi\rightarrow1$). Using \cite{abramo} 
\begin{align}
f_{1}(\xi) & =F\left( a,b,a+b+1-c;1-\xi\right) ,  \label{49} \\
f_{2}(\xi) & =(1-\xi)^{c-a-b}F\left( c-b,c-a,c-a-b+1;1-\xi\right) ,
\label{50}
\end{align}
we find in the limit $\xi\rightarrow1$, $f_{1}(\xi)\sim1$ and $f_{2}(\xi
)\sim(1-\xi)^{c-a-b}$. On the other hand, $(1-\xi)\sim p^{-2}$, so by
replacing $f_{1}(\xi)$ and $f_{2}(\xi)$ in Eq. (\ref{39}), we obtain the
following behavior of the two solutions 
\begin{align}
& \psi_{1}(p)\underset{p\rightarrow\infty}{\sim}p^{-2},  \label{51} \\
& \psi_{2}(p)\underset{p\rightarrow\infty}{\sim}p^{-3-2\omega_{4}}.
\label{52}
\end{align}
These two solutions can be found by considering the Schr\"{o}dinger equation
(\ref{31}) in the limit $p\rightarrow\infty$\ and seeking a solution in the
form $p^{s}$.

\bigskip This behavior is completely different from that of ordinary quantum
mechanics: both solutions are independent\ of the coupling constant;
moreover, the solution with asymptotic behavior (\ref{51}) does not depend
on the deformation parameters and falls off more slowly than $\psi_{2}$.
This implies that $\psi_{1}$ does not satisfy the boundary condition (\ref%
{31dd}), imposed by the integral equation, and so must be rejected. We
conclude that the physical wave function is $\psi_{2}$ with behavior at
infinity given by 
\begin{equation}
p^{2}\psi(p)\underset{p\rightarrow\infty}{\sim}p^{-1-2\omega_{4}}.
\label{53}
\end{equation}

The main conclusion, which we draw from this section, is that the singular
attractive $1/R^{2}$ potential is regularized by this minimal length, so
that the boundary condition (\ref{53}) will suffice to extract the energy
spectrum, as will be shown in Sec. IV D.

\subsection{ Full solution}

By the same technique as in the case $E=0$, Eq. (\ref{36}) can be rewritten
in a form of a known differential equation by choosing conveniently the
parameters $\lambda$ and $\lambda^{^{\prime}}$ of transformation (\ref{35}).
Taking $\lambda=1$ and $\lambda^{^{\prime}}=0$, Eq. (\ref{35}) reads 
\begin{equation}
\psi(z)=(1-z)f(z),  \label{58}
\end{equation}
and Eq. (\ref{36}) becomes after some calculations 
\begin{equation*}
\frac{d^{2}f(z)}{dz^{2}}+\left[ \frac{2}{(z-1)}+\dfrac{8\left[
(1+\omega)+(1-\omega)z\right] }{(2\omega-1)(z^{2}-1)(z-z_{0})}+\frac {%
(\omega_{2}z+\omega_{3})}{\omega_{1}(z^{2}-1)}\right] \frac{df(z)}{dz}
\end{equation*}%
\begin{equation}
+\left[ \dfrac{(2-\omega_{4}+\frac{\kappa}{1-2\omega})z-(1+\omega_{4}+\frac{%
\kappa}{1-2\omega})}{(z+1)(z-1)(z-z_{0})}\right] f(z)=0,  \label{59}
\end{equation}
with the notations defined by Eq. (\ref{33}), and :%
\begin{equation*}
z_{0}=\frac{2\omega+1}{2\omega-1}.
\end{equation*}

Equation (\ref{59}) is a linear homogeneous second-order differential
equation with four singularities $z=-1,1,z_{0},\infty$, all regular. So, Eq.
(\ref{59}) belongs to the class of Fuchsian equations, and can be
transformed into the canonical form of Heun's equation, having regular
singularities at $z=0,1,\xi_{0},\infty$ \cite{snow,ronveau}. The simple
change of variable 
\begin{equation*}
\xi=\frac{z+1}{2}
\end{equation*}
leads to the following canonical form of Heun's equation:%
\begin{equation}
\frac{d^{2}f(\xi)}{d\xi^{2}}+\left( \frac{c}{\xi}+\frac{e}{\xi-1}+\frac {d}{%
\xi-\xi_{0}}\right) \frac{df(\xi)}{d\xi}+\left( \frac{ab\xi+q}{\xi
(\xi-1)(\xi-\xi_{0})}\right) f(\xi)=0,  \label{60}
\end{equation}
with the parameters 
\begin{align}
a & =\frac{1}{2}(3-\omega_{4}-\widetilde{\nu}),\text{ \ \ \ \ }\widetilde{\nu%
}=\left[ (\omega_{4}-1)^{2}-\frac{4\kappa}{1-2\omega}\right] ^{\frac{1}{2}},
\notag \\
b & =\frac{1}{2}(3-\omega_{4}+\widetilde{\nu}),\text{ \ \ \ }\xi_{0}=\frac{%
2\omega}{2\omega-1},  \notag \\
c & =\frac{3}{2},\text{ \ \ \ \ }d=2,\text{ \ \ \ \ }e=\frac{1}{2}%
-\omega_{4},  \label{61} \\
q & =-\left( \frac{3}{2}+\frac{\kappa}{1-2\omega}\right) ,  \notag
\end{align}
which are linked by the Fuchsian condition 
\begin{equation}
a+b+1=c+d+e.  \label{62}
\end{equation}

In the neighborhood of $\xi=0$, the two linearly independent solutions of
Eq. (\ref{60}) are \cite{snow} 
\begin{equation}
f_{1}(\xi)=H(\xi_{0},q,a,b,c,d;\xi),  \label{63}
\end{equation}%
\begin{equation}
f_{2}(\xi)=\xi^{1-c}H(\xi_{0},q^{^{\prime}},1+a-c,1+b-c,2-c,d;\xi),
\label{64}
\end{equation}
where 
\begin{equation*}
q^{^{\prime}}=q-(1-c)\left[ d+\xi_{0}(1+a+b-c-d)\right] .
\end{equation*}
$H(\xi_{0},q,a,b,c,d;\xi)$ is the Heun function defined by the series 
\begin{equation}
H(\xi_{0},q,a,b,c,d;\xi)=1-\frac{q}{c\xi_{0}}\xi+\sum\limits_{n=2}^{\infty
}C_{n}\xi^{n},  \label{65}
\end{equation}
where the coefficients $C_{n}$\ are determined by the difference equation :%
\begin{equation*}
(n+2)(n+1+c)\xi_{0}C_{n+2}=\left\{ (n+1)^{2}(\xi_{0}+1)+(n+1)\left[
c+d-1\right. \right.
\end{equation*}%
\begin{equation}
\left. +\left. (a+b-d)\xi_{0}\right] -q\right\} C_{n+1}-(n+a)(n+b)C_{n},
\label{66}
\end{equation}
with the initial conditions 
\begin{equation*}
C_{0}=1,\ \text{\ }C_{1}=\frac{-q}{c\xi_{0}},\text{ and }C_{n}=0,\text{ if }%
n<0.
\end{equation*}

Now, we can write the full solution of the deformed Schr\"{o}dinger equation
(\ref{34}). Thus, by using Eq. (\ref{58}) the solution $\psi(\xi)$, which is
regular (finite) in the neighborhood of $\xi=0$, is given by 
\begin{equation}
\psi(\xi)=A(1-\xi)H(\xi_{0},q,a,b,c,d;\xi),  \label{67}
\end{equation}
where $A$ is a normalization constant.

We show in the Appendix that in the limit $\beta,\beta^{^{\prime}}\ll1$, we
recover the result of ordinary quantum mechanics, given by Eq. (5); in the
limit $E\rightarrow0$, the zero energy solution (\ref{46}) is obtained and
finally, as was expected, Eq. (\ref{67}) has the same behavior as Eq. (\ref%
{46}) in the limit $p\rightarrow\infty$.

\subsection{Eigenvalue problem}

\bigskip We now study in more detail the solution to Eq. (\ref{60}), to show
how the introduction of a minimal length regularizes the singular attractive 
$1/R^{2}$ potential. For this purpose, we begin by the special case $%
\beta=\beta^{^{\prime}}$.

\subsubsection{Special case $\protect\beta=\protect\beta^{^{\prime}}$}

In this case, the Heun equation (\ref{60}) is reduced to a hypergeometric
equation. Indeed, we have $\omega_{4}=\frac{1}{2}$, and hence 
\begin{align*}
e & =0, \\
ab & =-q=\frac{3}{2}+\frac{\kappa}{1-2\omega},
\end{align*}
and the Fuchsian condition (\ref{62}) becomes 
\begin{equation*}
a+b+1=c+d.
\end{equation*}

Using the change of variable 
\begin{equation*}
x=\frac{\xi}{\xi_{0}},
\end{equation*}
Eq. (\ref{60}) takes the form of a hypergeometric differential equation \cite%
{abramo}\ 
\begin{equation}
x(1-x)f^{^{\prime\prime}}(x)+\left[ c-(a+b+1)x\right] f^{^{%
\prime}}(x)-abf(x)=0  \label{68}
\end{equation}
with the parameters 
\begin{align}
a & =\frac{5}{4}-\frac{\widetilde{\nu}}{2},  \notag \\
b & =\frac{5}{4}+\frac{\widetilde{\nu}}{2},  \label{69} \\
c & =\frac{3}{2},\text{ \ \ }\widetilde{\nu}=\left[ \frac{1}{4}-\frac{4\kappa%
}{1-2\omega}\right] ^{\frac{1}{2}}.  \notag
\end{align}
The solution to the Schr\"{o}dinger equation, which is finite in the
vicinity of $\xi=0$, is 
\begin{equation}
\psi_{\beta=\beta^{^{\prime}}}(\xi)=A(1-\xi)F(a,b,c;\xi/\xi_{0}).  \label{70}
\end{equation}

\subsubsection{Energy spectrum}

To compute the energy spectrum, we merely require that the wave function (%
\ref{70}) satisfies the boundary condition (\ref{31dd}). Since 
\begin{align*}
1-\xi & =\frac{1}{1+2\beta p^{2}}\underset{p\rightarrow\infty}{\sim}p^{-2},
\\
\frac{\xi}{\xi_{0}} & =\frac{2\omega-1}{2\omega}\frac{\omega_{1}p^{2}}{%
1+\omega_{1}p^{2}}\underset{p\rightarrow\infty}{\sim}\frac{2\omega -1}{%
2\omega},
\end{align*}
the wave function (\ref{70}) behaves like 
\begin{equation*}
\psi_{\beta=\beta^{^{\prime}}}\underset{p\rightarrow\infty}{\sim}%
p^{-2}F(a,b,c;\frac{2\omega-1}{2\omega})\text{.}
\end{equation*}
From the boundary condition : $p^{2}\psi\underset{p\rightarrow\infty }{%
\rightarrow}0$, we then obtain the following condition: 
\begin{equation}
F(a,b,c;\frac{2\omega-1}{2\omega})=0.  \label{71}
\end{equation}

This equation constitutes the quantization condition; the eigenvalues $%
\omega $ are the zeros of the hypergeometric function.

Let us now consider the limit $\omega \equiv -2m\beta E\ll 1$, i.e., 
\begin{equation*}
\left\vert \frac{2\omega -1}{2\omega }\right\vert \sim \frac{1}{2\omega }\gg
1.
\end{equation*}%
By means of the transformation (\ref{7}), and by taking into account that $%
F(a,b,c;-2\omega )\underset{\omega \ll 1}{\approx }1$, Eq. (\ref{71}) can be
written in the following form :%
\begin{equation}
\omega ^{5/4}\left\{ \exp \left[ i\left( \arg [A]-\frac{\nu }{2}\ln [2\omega
]\right) \right] +\exp \left[ -i\left( \arg [A]-\frac{\nu }{2}\ln [2\omega
]\right) \right] \right\} =0,  \label{72}
\end{equation}%
where we have used the notations 
\begin{align*}
A& =\frac{\Gamma (i\nu )}{\Gamma (\frac{5}{4}+i\frac{\nu }{2})\Gamma (\frac{1%
}{4}+i\frac{\nu }{2})}=\left\vert A\right\vert \exp [i\arg (A)], \\
\nu & =\sqrt{4\kappa -1/4}.
\end{align*}%
From Eq. (\ref{72}), we have 
\begin{equation}
\cos \left[ \arg (A)-\frac{\nu }{2}\ln (2\omega )\right] =0,  \label{74}
\end{equation}%
which gives the following expression of the energy spectrum :%
\begin{equation}
E_{n}=\frac{-1}{4m\beta }\exp \left\{ \frac{2}{\nu }\left[ \arg (A)-(n+\frac{%
1}{2})\pi \right] \right\} ,  \notag
\end{equation}%
as one has%
\begin{equation}
\left\vert E_{n}\right\vert \ll \frac{1}{4m\beta }\text{ \ \ \ \ \ \ \ \ \ \
\ \ \ \ \ \ \ }n=0,1,2,...\text{ .}  \label{75}
\end{equation}

We recall that the deformation parameter $\beta$ is related to the minimal
length via Eq. (\ref{26}), hence $(\Delta r)_{\min}=2\hbar\sqrt{\beta}$.

The energy spectrum (\ref{75}) is identical to the one obtained by a cutoff
regularization [see Eq. (\ref{16})]. The parameter $\beta+\beta^{^{%
\prime}}=2\beta$ \ is simply the inverse square of the ultraviolet cutoff $%
\Lambda$.

Equation (\ref{75}) is accompanied by the condition $\left\vert
E_{n}\right\vert \ll1/4m\beta$, which excludes systematically the
undesirable values of the number $n$, so there is now a ground state with
finite energy. In the case of a weakly attractive potential ($4\kappa<1/4$),
Eq. (\ref{72}) has no solution.

These results are confirmed by the examination of the exact eigenvalue
equation (\ref{71}). We have plotted the hypergeometric function in Eq. (\ref%
{71}) as a function of $\omega=-2m\beta E$ for fixed $\kappa
=m\alpha/2\hbar^{2}$. The energy eigenvalues are the zeros of the function;
Figs. 1 and 2 show that the energy of the ground state ($\omega_{1}$) is
finite; for $\kappa=3/4$, $\omega_{1}\approx0.07$ and for $\kappa=2$, $%
\omega_{1}\approx0.37$. As in ordinary quantum mechanics, there are many,
almost identical, excited states with $\omega\simeq0$ (accumulation point).
The energy levels increase as we increase the coupling constant. In Fig. 2,
we can see the energy of the first excited state. Figure 3 shows that there
are no bound states for $\kappa=1/20$; we find that a critical coupling
constant $\kappa^{\ast}$, below which there are no bound states, has the
same value as in ordinary quantum mechanics, i.e., $\kappa^{\ast}=1/16$.

\begin{figure}[h]
\begin{center}
\includegraphics[width=7cm,height=5cm]{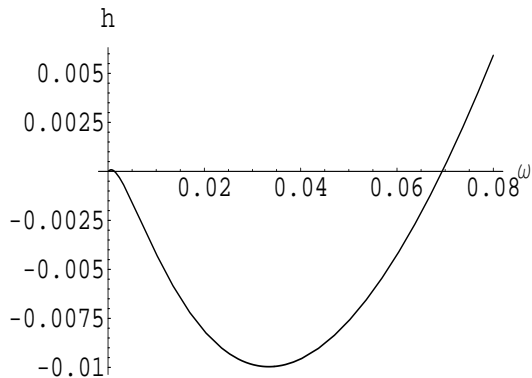}
\end{center}
\caption{$h\equiv F(a,b,c;\frac{2\protect\omega -1}{2\protect\omega })$ as a
function of $\protect\omega $, for $\protect\kappa =3/4.$ All quantities $a,$
$b,$ $c,$ $\protect\omega ,$ $\protect\kappa $ are dimensionless.}
\label{Fig. 1}
\end{figure}

\begin{figure}[h]
\begin{center}
\includegraphics[width=7cm,height=5cm]{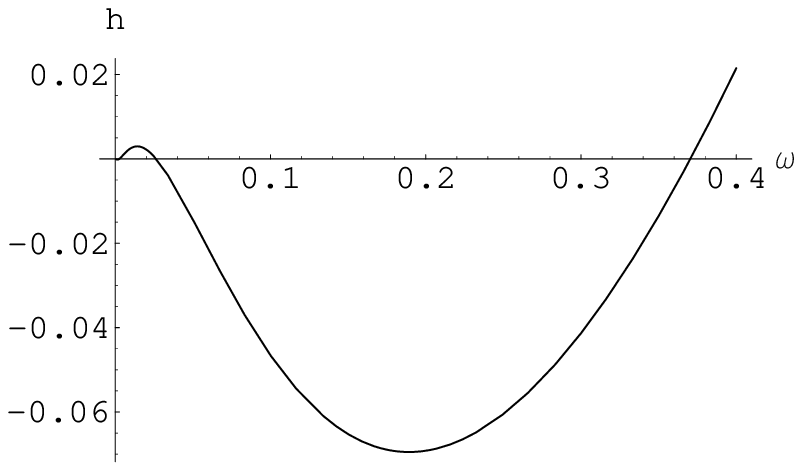}
\end{center}
\caption{$h\equiv F(a,b,c;\frac{2\protect\omega -1}{2\protect\omega })$ as a
function of $\protect\omega $, for $\protect\kappa =2.$ All quantities $a,$ $%
b,$ $c,$ $\protect\omega ,$ $\protect\kappa $ are dimensionless.}
\label{Fig. 2}
\end{figure}

\begin{figure}[h]
\begin{center}
\includegraphics[width=7cm,height=5cm]{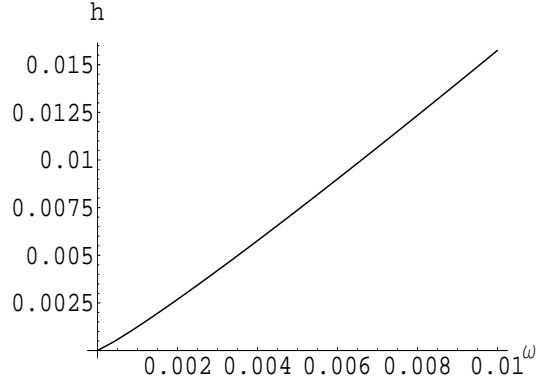}
\end{center}
\caption{$h\equiv F(a,b,c;\frac{2\protect\omega -1}{2\protect\omega })$ as a
function of $\protect\omega $, for $\protect\kappa =1/20.$ All quantities $%
a, $ $b,$ $c,$ $\protect\omega ,$ $\protect\kappa $ are dimensionless.}
\label{Fig. 3}
\end{figure}

An interesting feature of the expression of the energy (\ref{75}) is that it
is inversely proportional to the deformation parameter $\beta$; thus if $%
\beta$ is a very small parameter the energy of the ground state is very
large. Consequently, in the case of the inverse square potential, the
minimal length could be viewed as an intrinsic dimension of a system, as
argued by Kempf (see, for instance, \cite{k7}). However, if this minimal
length is obtained from calculations connected with the harmonic oscillator
and the hydrogen atom, as in \cite{chang,ben}, namely $\sim0.1$ fm, the
energy of the ground state would be so large, and thus it would not be in
the energy scale where nonrelativistic quantum mechanics is valid.

\subsubsection{Generalization to the case $\protect\beta\neq\protect\beta%
^{^{\prime}}$}

Let us return to the general solution (\ref{67}) 
\begin{equation*}
\psi(\xi)=A(1-\xi)H(\xi_{0},q,a,b,c,d;\xi).
\end{equation*}
It can be written in the form \cite{snow}\ 
\begin{equation}
\psi(\xi)=A(1-\xi)H(\frac{1}{\xi_{0}},\frac{q}{\xi_{0}},a,b,c,e;\frac{\xi}{%
\xi_{0}}).  \label{77}
\end{equation}

\bigskip As in the case $\beta=\beta^{^{\prime}}$ we impose the boundary
condition (\ref{31dd}), and obtain the following quantization condition: 
\begin{equation}
H(\frac{2\omega-1}{2\omega},\frac{2\omega-1}{2\omega}q,a,b,c,e;\frac {%
2\omega-1}{2\omega})=0.  \label{78}
\end{equation}

In the case where $\omega\ll1$, we set 
\begin{equation*}
\sigma=\frac{2\omega-1}{2\omega}\approx\frac{-1}{2\omega}\rightarrow\infty
\end{equation*}
by means of the following transformation \cite{snow}: 
\begin{equation}
H(\sigma,\sigma q,a,b,c,e;\xi)\underset{\sigma\rightarrow\infty}{=}F(\delta+%
\sqrt{\delta^{2}+q},\delta-\sqrt{\delta^{2}+q},c;\xi),\text{ }c\neq0,-1,-2,..%
\text{ ,}  \label{79}
\end{equation}
where%
\begin{equation*}
\text{ }\delta=\frac{a+b-e}{2}.
\end{equation*}
equation (\ref{78}) reads 
\begin{equation}
F(\frac{5}{4}-i\frac{\nu}{2},\frac{5}{4}+i\frac{\nu}{2},\frac{3}{2};\frac {-1%
}{2\omega})\underset{\omega=-(\beta+\beta^{^{\prime}})mE\ll1}{=}0.
\label{80}
\end{equation}

Obviously, we get the same expression of the energy spectrum as in the case $%
\beta=\beta^{^{\prime}}$. It is sufficient to replace in Eq. (\ref{75}), $%
2\beta$ by $\beta+\beta^{^{\prime}}$.

\section{Summary and conclusion}

We have solved exactly the problem of the singular inverse square potential
in the framework of quantum mechanics with a generalized uncertainty
relation implying the existence of a minimal length. In the momentum
representation, the wave function is a Heun function, which reduces to a
hypergeometric function for $E=0$ and for $\beta=\beta^{^{\prime}}$. The
potential is regularized in a natural way by this minimal length, so that
the energy spectrum is bounded from below. The results of ordinary quantum
mechanics with a regularizing cutoff ($\Lambda$) are recovered in the limit $%
\beta ,\beta^{^{\prime}}\ll1$; the parameter $\beta+\beta^{^{\prime}}$ \
plays the role of the inverse square of $\Lambda$.

In conclusion, this study shows that the idea of the introduction of a
minimal length, first proposed in high energy physics, could also apply to
nonrelativistic quantum mechanics. In the new formalism based on the
deformed Heisenberg algebra, the treatment of the singular $1/R^{2}$
potential is similar to that of regular potentials: we do not need to
introduce any arbitrary parameters because $\beta$ and $\beta^{^{\prime}}$
are physical parameters of the formalism, and describe the short distance
behavior of the interaction. The formalism includes a natural
\textquotedblright cutoff\textquotedblright\ and modifies the potential at
short distances, so that the energy spectrum is computed without imposing
any extra condition. The latter result leads us to conclude with Kempf \cite%
{k1,k7} that this elementary length should rather be viewed as an intrinsic
dimension of a system, at least for the problem considered here.

\appendix

\section{Limit $\protect\beta,\protect\beta^{^{\prime}}\ll1$}

We write the wave function in the form given by Eq. (\ref{77})%
\begin{equation}
\psi(\xi)=A(1-\xi)H(\frac{1}{\xi_{0}},\frac{q}{\xi_{0}},a,b,c,e;\frac{\xi}{%
\xi_{0}}),  \label{81}
\end{equation}
In the limit $\beta,\beta^{%
{\acute{}}%
}\ll1$, we have 
\begin{align*}
\xi & =\frac{\omega_{1}p^{2}}{1+\omega_{1}p^{2}}\approx\omega_{1}p^{2}\text{%
, where :\ }\omega_{1}=(\beta+\beta^{^{\prime}}), \\
\xi_{0} & =\frac{2\omega}{2\omega-1}\approx-2\omega,\text{ \ where : }%
\omega=-m\omega_{1}E, \\
\frac{\xi}{\xi_{0}} & \approx\frac{p^{2}}{2mE}\text{ \ \ and \ }1-\xi
\approx1,
\end{align*}
hence 
\begin{equation}
\psi(y)\underset{\beta,\beta^{%
{\acute{}}%
}\ll1}{\approx}H(\sigma,\sigma\widetilde{q},\widetilde{a},\widetilde {b}%
,c,e;y),  \label{82}
\end{equation}
where we have used the notations $\sigma=\frac{1}{2m\omega_{1}E}$, \ $y=%
\frac{p^{2}}{2mE}$, and $\widetilde{a},$ $\widetilde{b}$, $\widetilde{q}$
are the limits of the parameters $a$, $b$, $q$ when $\beta,\beta^{^{\prime}}%
\ll1$.

By means of the transformation (\ref{79}), Heun's function is transformed to
a hypergeometric function, given by 
\begin{equation}
H(\sigma,\sigma\widetilde{q},\widetilde{a},\widetilde{b},c,e;y)\underset{%
\sigma\rightarrow\infty}{=}F(\delta+\sqrt{\delta^{2}+\widetilde{q}},\delta-%
\sqrt{\delta^{2}+\widetilde{q}},c;y),\text{ }c\neq0,-1,-2,..\text{ ,}
\label{83}
\end{equation}
where 
\begin{equation*}
\text{ }\delta=\frac{\widetilde{a}+\widetilde{b}-e}{2}.
\end{equation*}
After a direct calculation we get 
\begin{equation*}
\psi(p)\underset{\beta,\beta^{^{\prime}}\ll1}{\approx}F(\frac{5}{4}+i\frac {%
\nu}{2},\frac{5}{4}-i\frac{\nu}{2},\frac{3}{2};\frac{p^{2}}{2mE}).
\end{equation*}

It is exactly the wave function in momentum representation for the
attractive $-\alpha/R^{2}$ potential in ordinary quantum mechanics [see Eq. (%
\ref{5})] .

\section{Limit $p\rightarrow\infty$}

To examine the behavior of $\psi(\xi)$, when $p\rightarrow\infty$ ($%
\xi\rightarrow1$), we use the well-known relation \cite{snow,chet} 
\begin{align}
H(\xi_{0},q,a,b,c,d;\xi) & =C_{1}H(1-\xi_{0},-q-ab,a,b,e,d;1-\xi)  \notag \\
& +C_{2}(1-\xi)^{1-e}H(1-\xi_{0},q_{2},c+d-a,c+d-b,2-e,d;1-\xi),  \label{84}
\end{align}
where 
\begin{align*}
C_{1} & =H(\xi_{0},q,a,b,c,d;1), \\
C_{2} & =H(\xi_{0},q-\xi_{0}c[1-e],c+d-a,c+d-b,c,d;1), \\
q_{2} & =-q-ab-(1-e)[d+c(1-\xi_{0})].
\end{align*}
By adopting the Heun normalization $H(\xi_{0},q,a,b,c,d;0)=1$, the wave
function (\ref{67}), in the limit $p\rightarrow\infty$, behaves as follows:%
\begin{equation*}
\psi(\xi)\underset{\xi\rightarrow1}{\approx}C_{1}(1-\xi)+C_{2}(1-\xi)^{2-e},
\end{equation*}
and since 
\begin{equation*}
1-\xi\underset{p\rightarrow\infty}{\approx}p^{-2}\text{, \ \ \ }2-e=\frac {3%
}{2}+\omega_{4},
\end{equation*}
then, the asymptotic behavior of $\psi(p)$ in this region is 
\begin{equation}
\psi(p)\underset{p\rightarrow\infty}{\approx}C_{1}p^{-2}+C_{2}p^{-(3+2\omega
_{4})}.  \label{85}
\end{equation}

This behavior is identical to that of the zero energy solution [see Eqs (\ref%
{51}) and (\ref{52})] because Schr\"{o}dinger equation does not depend on
the energy in the limit $p\rightarrow\infty.$

\section{\protect\bigskip Limit $E\rightarrow0$}

We show, here, that the zero energy solution (\ref{46}) can be obtained from
the full solution (\ref{67}) in the limit $E\rightarrow0$. For this purpose,
let us return to the transformation (\ref{84}). By taking into acount that $%
\xi_{0}\rightarrow0$ and $\omega\rightarrow0$ when $E\rightarrow0$, the wave
function (\ref{67}) can be written as 
\begin{align}
& \psi(\xi)\underset{E\rightarrow0}{=}A(1-\xi)\left[
C_{1}H(1,q_{1},a_{1},b_{1},c_{1},d_{1};1-\xi)\right.  \notag \\
& \left. +C_{2}(1-\xi)^{\frac{1}{2}%
+\omega_{4}}H(1,q_{2},a_{2},b_{2},c_{2},d_{2};1-\xi)\right] ,  \label{86}
\end{align}
with the parameters

\begin{equation*}
\begin{array}{c}
q_{1}=-\frac{1}{2}+\omega_{4}, \\ 
a_{1}=\frac{1}{2}(3-\omega_{4}-\widetilde{\nu}_{0}), \\ 
b_{1}=\frac{1}{2}(3-\omega_{4}+\widetilde{\nu}_{0}), \\ 
c_{1}=\frac{1}{2}-\omega_{4}, \\ 
d_{1}=2,%
\end{array}
\ \ \ \ \ \ \ \ 
\begin{array}{c}
q_{2}=-\frac{9}{4}+\frac{5\omega_{4}}{2}, \\ 
\text{\ }a_{2}=2+\frac{\omega_{4}}{2}+\frac{\widetilde{\nu}_{0}}{2}, \\ 
b_{2}=2+\frac{\omega_{4}}{2}-\frac{\widetilde{\nu}_{0}}{2}, \\ 
c_{2}=\frac{3}{2}+\omega_{4}, \\ 
d_{2}=2,%
\end{array}%
\end{equation*}
where 
\begin{equation*}
\text{\ }\widetilde{\nu}_{0}\equiv\widetilde{\nu}(\omega=0)=\sqrt{(\omega
_{4}-1)^{2}-4\kappa}.
\end{equation*}
We use once again another transformation of the Heun functions \cite{snow},%
\begin{align*}
H(1,q,a,b,c,d;\xi) & =(1-\xi)^{\frac{c-a-b}{2}+\tau}F(\frac{c+a-b}{2}+\tau,%
\frac{c-a+b}{2}+\tau,c;\xi),\text{ } \\
\text{where } & \text{: }\tau=\pm\sqrt{(\frac{c-a-b}{2})^{2}-ab-q},\text{\ \
if }c\neq0,-1,-2,..
\end{align*}
For the two Heun's functions in Eq. (\ref{86}), a direct calculation gives
the following result :%
\begin{equation*}
\tau_{1}=\tau_{2}=\pm\sqrt{\frac{1}{16}-\kappa}=\pm i\frac{\nu}{2}
\end{equation*}
By choosing, for convenience, the sign ($-$), the wave function (\ref{86})
reads as follows: 
\begin{align}
& \psi(\xi)\underset{E\rightarrow0}{=}A(1-\xi)\xi^{-\frac{5}{4}-i\frac{\nu }{%
2}}\left[ C_{1}F(\widetilde{a}_{1},\widetilde{b}_{1},\widetilde{c}%
_{1};1-\xi)\right.  \notag \\
& \left. +C_{2}(1-\xi)^{\frac{1}{2}+\omega_{4}}F(\widetilde{a}_{2},%
\widetilde{b}_{2},\widetilde{c}_{2};1-\xi)\right] ,  \label{87}
\end{align}
with the following parameters:%
\begin{equation*}
\begin{array}{c}
\widetilde{a}_{1}=\frac{1}{4}-\frac{\omega_{4}}{2}-\frac{\mu}{2}-i\frac{\nu 
}{2}, \\ 
\widetilde{b}_{1}=\frac{1}{4}-\frac{\omega_{4}}{2}+\frac{\mu}{2}-i\frac{\nu 
}{2}, \\ 
\widetilde{c}_{1}=\frac{1}{2}-\omega_{4},%
\end{array}
\ \ \ \ \ \ \ \ 
\begin{array}{c}
\widetilde{a}_{2}=\frac{3}{4}+\frac{\omega_{4}}{2}+\frac{\mu}{2}-i\frac{\nu 
}{2}, \\ 
\text{\ }\widetilde{b}_{2}=\frac{3}{4}+\frac{\omega_{4}}{2}-\frac{\mu}{2}-i%
\frac{\nu}{2}, \\ 
\widetilde{c}_{2}=\frac{3}{2}+\omega_{4},%
\end{array}%
\end{equation*}
\bigskip where 
\begin{equation*}
\mu=\sqrt{(\omega_{4}-1)^{2}-4\kappa}\text{, \ \ }\kappa=m\alpha/2\hbar^{2}.
\end{equation*}
It is easily seen that the expression between brackets in Eq. (\ref{87}) is
exactly a linear combination of the two solutions (\ref{49}) and (\ref{50}),
in the vicinity of $\xi=1$, of the hypergeometric equation (\ref{42}), with
the parameters $a,b,c$, given by \cite{abramo} 
\begin{equation*}
\begin{array}{c}
a=\widetilde{a}_{1}, \\ 
b=\widetilde{b}_{1}, \\ 
c=1-i\nu.%
\end{array}%
\end{equation*}
Obviously, in the neighborhood of $\xi=0$, we have the solution (\ref{46}).

\bigskip

\begin{acknowledgments}
D. B thanks Professor Tahar Boudjedaa for several very instructive
discussions, especially concerning Heun's differential equations, and
acknowledges the Belgian Technical Cooperation ( BTC) and the Algerian
ministry of Higher Education and Scientific Research (MESRS) for their
financial support.The work of M. B was supported by the National Fund for
Scientific Research (FNRS), Belgium.
\end{acknowledgments}

\end{document}